\documentclass[12pt]{article}
\usepackage[utf8]{inputenc}
\usepackage{amsmath}
\usepackage{amssymb}
\usepackage{bbold}
\usepackage[title]{appendix}
\usepackage{comment}
\usepackage{float}
\usepackage{array}
\usepackage{hyperref}
\usepackage{natbib}
\usepackage{graphicx}
\usepackage{xcolor}

\newcommand{\newc}{\newcommand}
\newc{\bc}{\begin{center}}
\newc{\ec}{\end{center}}
\newc{\be}{\begin{enumerate}}
\newc{\ee}{\end{enumerate}}
\newc{\bi}{\begin{itemize}}
\newc{\ei}{\end{itemize}}
\newc{\bd}{\begin{description}}
\newc{\ed}{\end{description}}
\newc{\und}[1]{\underline{#1}}
\newc{\E}{\mbox{E}}
\newc{\V}{\mbox{V}}
\newc{\N}{\mbox{N}}
\newc{\B}{\mbox{Bin}}
\newc{\Bern}{\mbox{Bern}}
\newc{\Po}{\mbox{Po}}
\newc{\IG}{\mbox{IG}}
\newc{\Gam}{\mbox{Gam}}
\newc{\bdP}{\mbox{P}}
\newc{\bdp}{\mathsf{p}}
\newc{\bdphat}{\hat{\mathsf{p}}}
\newc{\odds}{\mbox{odds}}
\newc{\OR}{\mbox{OR}}
\newc{\stderr}{\mbox{s.e.}}
\newc{\logit}{\mbox{logit}}
\newc{\sign}{\mbox{sign}}
\newc{\SD}{\mbox{SD}}
\newc{\bdmu}{\mbox{\boldmath $\mu$}}
\newc{\bdSigma}{\mbox{\boldmath $\Sigma$}}
\newc{\bdLambda}{\mbox{\boldmath $\Lambda$}}
\newc{\bdmuhat}{\mbox{\boldmath $\hat{\mu}$}}
\newc{\bdeta}{\mbox{\boldmath $\eta$}}
\newc{\bdtheta}{\mbox{\boldmath $\theta$}}
\newc{\bdbeta}{\mbox{\boldmath $\beta$}}
\newc{\bdgamma}{\mbox{\boldmath $\gamma$}}
\newc{\bdetahat}{\mbox{\boldmath $\hat{\eta}$}}
\newc{\bdbetahat}{\mbox{\boldmath $\hat{\beta}$}}
\newc{\bdgammahat}{\mbox{\boldmath $\hat{\gamma}$}}
\newc{\bdthetahat}{\mbox{\boldmath $\hat{\theta}$}}
\newc{\bdvareps}{\mbox{\boldmath $\varepsilon$}}
\newc{\bdzero}{\mbox{\boldmath $0$}}
\newc{\bdone}{\mbox{\boldmath $1$}}
\newc{\bdnu}{\mbox{\boldmath $\nu$}}
\newc{\bdell}{\mbox{\boldmath $\ell$}}
\newc{\bdxi}{\mbox{\boldmath $\xi$}}
\newc{\bdomega}{\mbox{\boldmath $\omega$}}
\newc{\bdepsilon}{\mbox{\boldmath $\varepsilon$}}
\newc{\bdI}{\mathbf{I}}
\newc{\bdX}{\mbox{\boldmath $X$}}
\newc{\bdA}{\mbox{\boldmath $A$}}
\newc{\bdB}{\mbox{\boldmath $B$}}
\newc{\bdC}{\mbox{\boldmath $C$}}
\newc{\bdD}{\mbox{\boldmath $D$}}
\newc{\bdG}{\mbox{\boldmath $G$}}
\newc{\bdJ}{\mbox{\boldmath $J$}}
\newc{\Ktil}{\tilde{K}}
\newc{\Khat}{\hat{K}}
\newc{\bda}{\mbox{\boldmath $a$}}
\newc{\bdb}{\mbox{\boldmath $b$}}
\newc{\bdc}{\mbox{\boldmath $c$}}
\newc{\bde}{\mbox{\boldmath $e$}}
\newc{\bdu}{\mbox{\boldmath $u$}}
\newc{\bdv}{\mbox{\boldmath $v$}}
\newc{\bdx}{\mbox{\boldmath $x$}}
\newc{\bdy}{\mbox{\boldmath $y$}}
\newc{\bdz}{\mbox{\boldmath $z$}}
\newc{\bdr}{\mbox{\boldmath $r$}}
\newc{\bdQ}{\mbox{\boldmath $Q$}}
\newc{\bdR}{\mbox{\boldmath $R$}}
\newc{\bdY}{\mbox{\boldmath $Y$}}
\newc{\bdT}{\mbox{\boldmath $T$}}
\newc{\bdW}{\mbox{\boldmath $W$}}
\newc{\bdWtil}{\tilde{\mbox{\boldmath $W$}}}
\newc{\bdH}{\mbox{\boldmath $H$}}
\newc{\bdL}{\mbox{\boldmath $L$}}
\newc{\bdU}{\mbox{\boldmath $U$}}
\newc{\bdV}{\mbox{\boldmath $V$}}
\newc{\Multinom}{\mbox{Multinom}}
\newc{\Var}{\mbox{Var}}
\newc{\var}{\mbox{var}}
\newc{\diag}{\mbox{diag}}
\newc{\thetahat}{\hat{\theta}}
\newc{\tr}{\mbox{tr}}
\newc{\phat}{\hat{p}}
\newc{\Xbar}{\bar{X}}
\newc{\xbar}{\bar{x}}
\newc{\Ybar}{\bar{Y}}
\newc{\ybar}{\bar{y}}
\newc{\dbar}{\bar{d}}
\newc{\yhat}{\hat{y}}
\newc{\bdyhat}{\mbox{\boldmath $\hat{y}$}}
\newc{\ytil}{\tilde{y}}
\newc{\ftil}{\tilde{f}}
\newc{\Ho}{\mbox{\bf H}_o}
\newc{\Ha}{\mbox{\bf H}_a}
\newc{\phatYX}{\phat_Y - \phat_X}
\newc{\SSG}{\mbox{SSG}}
\newc{\SSB}{\mbox{SSB}}
\newc{\SSE}{\mbox{SSE}}
\newc{\SST}{\mbox{SST}}
\newc{\SSR}{\mbox{SSR}}
\newc{\SSAB}{\mbox{SSAB}}
\newc{\MSG}{\mbox{MSG}}
\newc{\MSB}{\mbox{MSB}}
\newc{\MSE}{\mbox{MSE}}
\newc{\MST}{\mbox{MST}}
\newc{\MSAB}{\mbox{MSAB}}
\newc{\dfE}{\mbox{dfE}}
\newc{\dfG}{\mbox{dfG}}
\newc{\dfB}{\mbox{dfB}}
\newc{\dfT}{\mbox{dfT}}
\newc{\dfAB}{\mbox{dfAB}}
\newc{\muhat}{\hat{\mu}}
\newc{\betahat}{\hat{\beta}}
\newc{\alphahat}{\hat{\alpha}}
\newc{\etahat}{\hat{\eta}}
\newc{\phihat}{\hat{\phi}}
\newc{\sigmahat}{\hat{\sigma}}
\newc{\cl}{\centerline}
\newc{\R}{\mathbb{R}}
\newc{\trans}{^\mathsf{T}}
\newc{\xtx}{\bdX\trans\bdX}
\newc{\xxtxx}{\bdX(\xtx)^{-1}\bdX\trans}
\newc{\argmin}{\operatornamewithlimits{argmin}}
\newc{\argmax}{\operatornamewithlimits{argmax}}
\newc{\incg}[2]{
\includegraphics[width=#1\textwidth]{#2}
}

\title{Paired comparison models with strength-dependent ties and order effects}
\author{Mark E.\ Glickman\thanks{Address for correspondence: 
Department of Statistics, Harvard University, 1 Oxford Street,
Cambridge, MA  02138.  Email address: {\tt glickman@fas.harvard.edu}} \\ 
Department of Statistics\\
Harvard University
}
\date{}

\begin{document}

\maketitle

\addtolength{\baselineskip}{12pt}

\begin{abstract}
Paired comparison models, such as the \citet{bradley_rank_1952} 
model and its
variants, are commonly used to measure competitor strength in games
and sports.  Extensions have been proposed to account for order effects
(e.g., home-field advantage) as well as the possibility of a tie as a
separate outcome, but such models are rarely adopted in practice due
to poor fit with actual data.  We propose a novel paired comparison model
that accounts not only for ties and order effects, but recognizes two
phenomena that are not addressed with commonly used models.  First,
the probability of a tie may be greater for stronger pairs of competitors.
Second, order effects may be more pronounced for stronger competitors.
This model is motivated in the context of tournament chess game outcomes.
The models are demonstrated on the results of US Chess Open game outcomes
from 2006 to 2019, large tournaments consisting of
players of wide-ranging strengths.
% , and to the Vienna 1898 chess tournament,
% a double-round robin tournament consisting of 20 of the world's top players.
\end{abstract}


\section{Introduction}\label{sec:intro}

Paired comparison models are widely used in the statistical analysis of competitive outcomes, 
particularly in sports and games, to estimate underlying competitor strength. 
In their simplest form, these models assign scalar strength parameters to 
each competitor,
and model the probability of one competitor defeating the other as a function 
of the difference in their strengths. 
Classic formulations such as the Bradley-Terrry model \citep{bradley_rank_1952} and the
Thurstone-Mosteller model \citep{thurstone1927law, mosteller_remarks_1951} capture binary outcomes (win/loss) 
but do not account for ties, which are prevalent in many contexts, 
including chess, boxing, and soccer. 
Ignoring ties can lead to model misspecification and loss of predictive accuracy, 
particularly when ties are frequent and non-random.

% Insert paragraph about background on draws.  Also clarify ``order effects.''
Despite the widespread occurrence of ties in real-world paired comparison settings, 
the development of models that explicitly accommodate tied outcomes has received 
relatively sparse attention in the statistical literature. 
One early and influential contribution is due to \citet{Davidson1970On},
who proposed a simple extension of the Bradley–Terry model to allow for ties 
by introducing a parameter that governs the probability of a draw. 
His model preserves the independence of irrelevant alternatives (IIA) 
and expresses tie likelihood as a function of the geometric mean of the 
two players’ strengths. 
\citet{davidson1977extending} extended this formulation to include order effects, 
such as home-field advantage or the benefit of playing white in chess, 
by incorporating an additive shift to the log-odds. 
More recent work by \citet{Baker2020Modifying}
offers a probabilistically 
grounded approach by using discrete distributions, most notably the 
geometric distribution, to derive modified versions of the Bradley–Terry and 
Plackett–Luce models that naturally permit tied outcomes. 
In a complementary line of work, \citet{Hankin2020}
proposed a generalized framework 
in which a draw is conceptualized as a third competitor, allowing the model 
to treat each match as a three-way contest and enabling the inclusion of both 
tie propensity and order effects within a unified likelihood-based system. 
While such efforts offer valuable tools for applied analysis, they remain 
relatively underexplored given the empirical importance of ties in many 
competitive contexts.

In this paper, we introduce a novel extension to standard paired comparison 
frameworks that  accounts for variation in both tie probability and order 
effects based on the average strength of the players. 
The motivation for this approach stems from empirical patterns observed in 
high-level competitive  settings, particularly in chess, where draws are 
more common between stronger players. 
This is plausibly explained by the tendency of stronger competitors to 
make fewer mistakes  and to be more effective at neutralizing each 
other's strategies, thereby increasing the  likelihood of a draw.
At the same time, the advantage of playing white in chess, commonly attributed 
to the  first-move initiative, is often more effectively exploited by stronger 
players, who are better  equipped to convert even small advantages into a win. 
These observations suggest that both the frequency of ties and the impact 
of positional  advantages are not static across the range of player strengths, 
but instead intensify among  more skilled competitors. 
Our proposed model captures this dynamic by allowing the probability of a draw 
and the advantage  of playing white to increase with the average ability of the 
two players involved in the match. 
This leads to a more realistic and flexible modeling framework that reflects 
the nuanced  nature of competitive interactions at different skill levels.

This paper is organized as follows.
We introduce traditional models with tie extensions for paired comparisons 
in Section~\ref{sec:background}, and demonstrate empirically the type
of deficiency exhibited in these models.
We propose our novel extension of these traditional models in
Section~\ref{sec:model} that accounts for draw probabilities and order
effects that depend on strength, 
and discuss model fitting in a likelihood-based and Bayesian setting.
We apply our approach in Section~\ref{sec:application} to 
% two  data sets;  one involving 
chess game outcomes from the US Chess Opens from 2006 to 2019.
% and the second fitting our model to the Vienna 1898 chess tournament,
% a double-round robin event involving 20 top players of the turn of the century.
The paper concludes in Section~\ref{sec:discuss}.

\section{Background and motivation}\label{sec:background}

The focus of our work is on extending the \citet{bradley_rank_1952} model,
a widely used framework for modeling outcomes in head-to-head competition,
for ties and order effects.
For the original Bradley-Terry model,
if two competitors $i$ and $j$ face each other, 
the probability that $i$ defeats $j$ is modeled as
\begin{equation}\label{eq:bt}
\text{logit} \, P(Y_{ij} = 1) = \theta_i - \theta_j,
\end{equation}
where $Y_{ij} = 1$ indicates a win for player $i$, and $\theta_i$ and $\theta_j$ represent the 
latent strength parameters of players $i$ and $j$, respectively. 
This model assumes only binary outcomes and does not account for the possibility of ties 
or contextual factors like order effects (e.g., home-field advantage, or white's advantage in chess).

To incorporate order effects, \citet{davidson1977extending} extended the Bradley-Terry model 
by including a covariate $x_{ij}$ that encodes the direction of the advantage.
In the context of chess,
we let
$x_{ij} = 1$ if player $i$ plays white, and $x_{ij} = -1$ if player $j$ plays white. 
The extended model can be written as
\begin{equation}\label{eq:db}
\text{logit} \, P(Y_{ij} = 1) = \theta_i - \theta_j + \frac{\alpha}{2} x_{ij},
\end{equation}
where $\alpha$ is the log-odds advantage conferred by playing white
in this parameterization.

Separately, ties can be incorporated by extending the outcome space to three categories,
in which a win ($Y_{ij} = 1$), loss ($Y_{ij} = 0$), and tie ($Y_{ij} = \frac{1}{2}$)
are the three possible outcomes.
A commonly used model that preserves Luce’s choice axiom, 
or the independence of irrelevant alternatives (IIA), 
is the model by \citet{Davidson1970On} % Davidson model \citep{davidson1970}, 
which assumes 
\begin{equation}
\label{eq:davidson}
\begin{aligned}
P(Y_{ij} = 1) &= \dfrac{\exp(\theta_i)}{\exp(\theta_i) + \exp(\theta_j) + \nu \cdot \exp\left( \frac{\theta_i + \theta_j}{2} \right)}, \\
P(Y_{ij} = 0) &= \dfrac{\exp(\theta_j)}{\exp(\theta_i) + \exp(\theta_j) + \nu \cdot \exp\left( \frac{\theta_i + \theta_j}{2} \right)}, \\
P(Y_{ij} = \tfrac{1}{2}) &= \dfrac{\nu \cdot \exp\left( \frac{\theta_i + \theta_j}{2} \right)}{\exp(\theta_i) + \exp(\theta_j) + \nu \cdot \exp\left( \frac{\theta_i + \theta_j}{2} \right)}.
\end{aligned}
\end{equation}
where $\nu > 0$ is a parameter that controls the overall likelihood of a tie. 
An appeal of this
model is that it remains invariant to the addition of a constant to 
both $\theta_i$ and $\theta_j$, thereby preserving the scale of the strength parameters.
The IIA property of~(\ref{eq:davidson}) implies that 
the odds of a win versus a loss between two competitors 
are independent of the probability of a tie. 
That is, the ratio $\bdP(\mbox{win})/\bdP(\mbox{loss})$ is unaffected by changes in the 
tie probability. 
This property ensures that introducing ties as a third outcome does not distort the 
model’s assessment of how the two competitors compare in terms of winning or losing. 
% Also, because this model preserves IIA, the ratio of the probability of a win to a loss
% does not change as a function of the parameter $\nu$.

Combining both ties and order effects into a single model was proposed by David (1988). 
David's model can be specified with outcome probabilities 
\begin{equation}\label{eq:david}
\begin{aligned}
P(Y_{ij} = 1) &= \dfrac{\exp\left(\theta_i + \frac{\alpha}{4} x_{ij}\right)}{\exp\left(\theta_i + \frac{\alpha}{4} x_{ij}\right) + \exp\left(\theta_j - \frac{\alpha}{4} x_{ij}\right) + \nu \cdot \exp\left(\frac{\theta_i + \theta_j}{2}\right)}, \\
P(Y_{ij} = 0) &= \dfrac{\exp\left(\theta_j - \frac{\alpha}{4} x_{ij}\right)}{\exp\left(\theta_i + \frac{\alpha}{4} x_{ij}\right) + \exp\left(\theta_j - \frac{\alpha}{4} x_{ij}\right) + \nu \cdot \exp\left(\frac{\theta_i + \theta_j}{2}\right)}, \\
P(Y_{ij} = \tfrac{1}{2}) &= \dfrac{\nu \cdot \exp\left(\frac{\theta_i + \theta_j}{2}\right)}{\exp\left(\theta_i + \frac{\alpha}{4} x_{ij}\right) + \exp\left(\theta_j - \frac{\alpha}{4} x_{ij}\right) + \nu \cdot \exp\left(\frac{\theta_i + \theta_j}{2}\right)}.
\end{aligned}
\end{equation}
The reason for the incorporation of $\alpha/4$ in the above expression is that,
% when $\nu=0$, the 
ignoring ties, the
log-odds of winning is given by
\begin{equation}\label{eq:alpha.over.4}
\begin{aligned}
\log\left( \frac{P(Y_{ij} = 1)}{P(Y_{ij} = 0)} \right)
&= \log\left( \frac{\exp\left(\theta_i + \frac{\alpha}{4} x_{ij}\right)}{\exp\left(\theta_j - \frac{\alpha}{4} x_{ij}\right)} \right) \\
&= \left( \theta_i + \frac{\alpha}{4} x_{ij} \right) - \left( \theta_j - \frac{\alpha}{4} x_{ij} \right) \\
&= \theta_i - \theta_j + \frac{\alpha}{2} x_{ij}
\end{aligned}
\end{equation}
which is the \citet{davidson1977extending} model as a special case.

While these Bradley-Terry extensions represent tractable methods to include the 
probability of a tie as a paired comparison outcome, 
they still exhibit important limitations. 
Most notably, they assume that the tie probability and the order effect are constant 
across all levels of player strength. 
One might argue that stronger players in head-to-head games tend to play less erratically
than weaker players, so that the probability of a tie can be expected to be higher
for stronger players.
Similarly, stronger players may be better at capitalizing on intrinsic advantages, such 
as the advantage in chess of playing white.

To see this outside a formal modeling framework, 
we performed an analysis of chess game results among tournament players who competed
in World Chess Federation (FIDE) events between 1995 and 2007.
The games appeared in Chess Informants ({\tt https://sahovski.com/}), 
tri-annual bulletins containing hundreds
of games from international tournaments by chess masters.
For each of the games in the data set, we recorded the FIDE ratings of the 
players involved and the game outcome (win, draw, loss) relative to the player
with the white pieces.
The FIDE rating system from 1995 to 2007, which implemented a variant of the system
developed by \citet{elo1978rating}, 
produced ratings that can be understood as linear transformations of Bradley-Terry 
parameters.
Specifically, if 
% $\hat{\theta}_i$
$\thetahat_i$
is an estimate of the Bradley-Terry parameter for player $i$, then 
$R_i = 1500 + \frac{400}{\log 10}\thetahat_i$ is the rating estimate on the scale
of FIDE/Elo ratings.
More discussion of the connection can be found in \citep{glickman2024models}.
We restricted our focus to games in which the rating difference was no more than 200
rating points, corresponding to an expected game outcome of 0.76.
This resulted in a total of 19,453 games.

\begin{figure}[ht]
\centerline{
\includegraphics[width=0.9\textwidth]{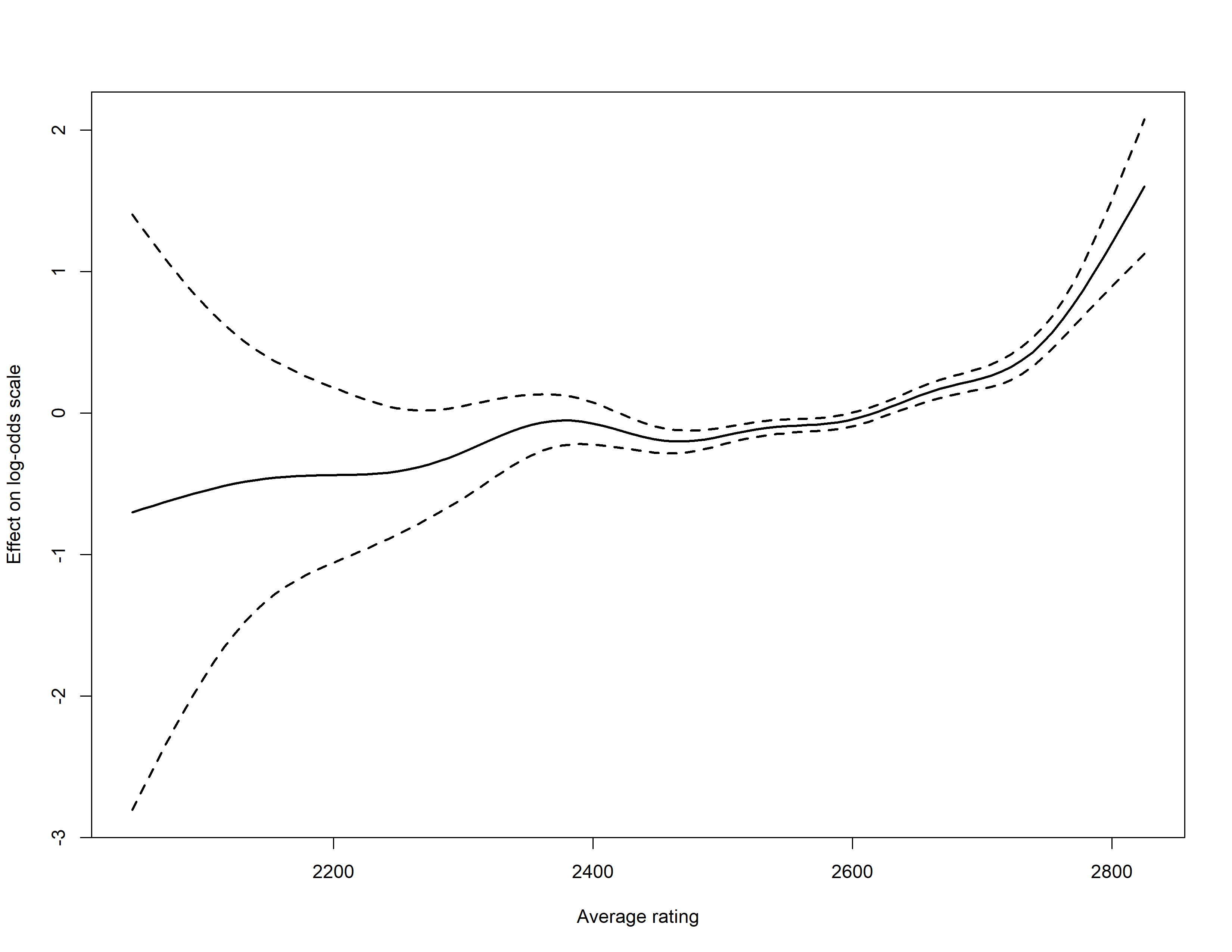}
}
\caption{Log-odds of white drawing as a function of the average rating between two 
players, controlling for rating difference.
The dashed lines are 95\% pointwise confidence limits.}
\label{fig:draw2}
\end{figure}

We first addressed the question of how the probability of a draw depends on 
within-pair average rating, controlling for the rating difference.
This was accomplished by fitting a generalized additive model \citep{hastie1986generalized}
with a logit link
for the probability a game resulted in a draw (versus a decisive result)
as the additive contribution of
a cubic smoothing spline function of the within-pair rating difference and
a cubic smoothing spline function of the within-pair average rating.
Figure~\ref{fig:draw2} displays the fitted smooth contribution of the 
within-pair average rating on the log-odds of a draw.
The figure clearly exhibits that
the higher the average rating of the two players, 
particularly for higher-rated players, 
the higher the log-odds of a draw, controlling for the rating difference.

\begin{figure}[ht]
\centerline{
\includegraphics[width=0.9\textwidth]{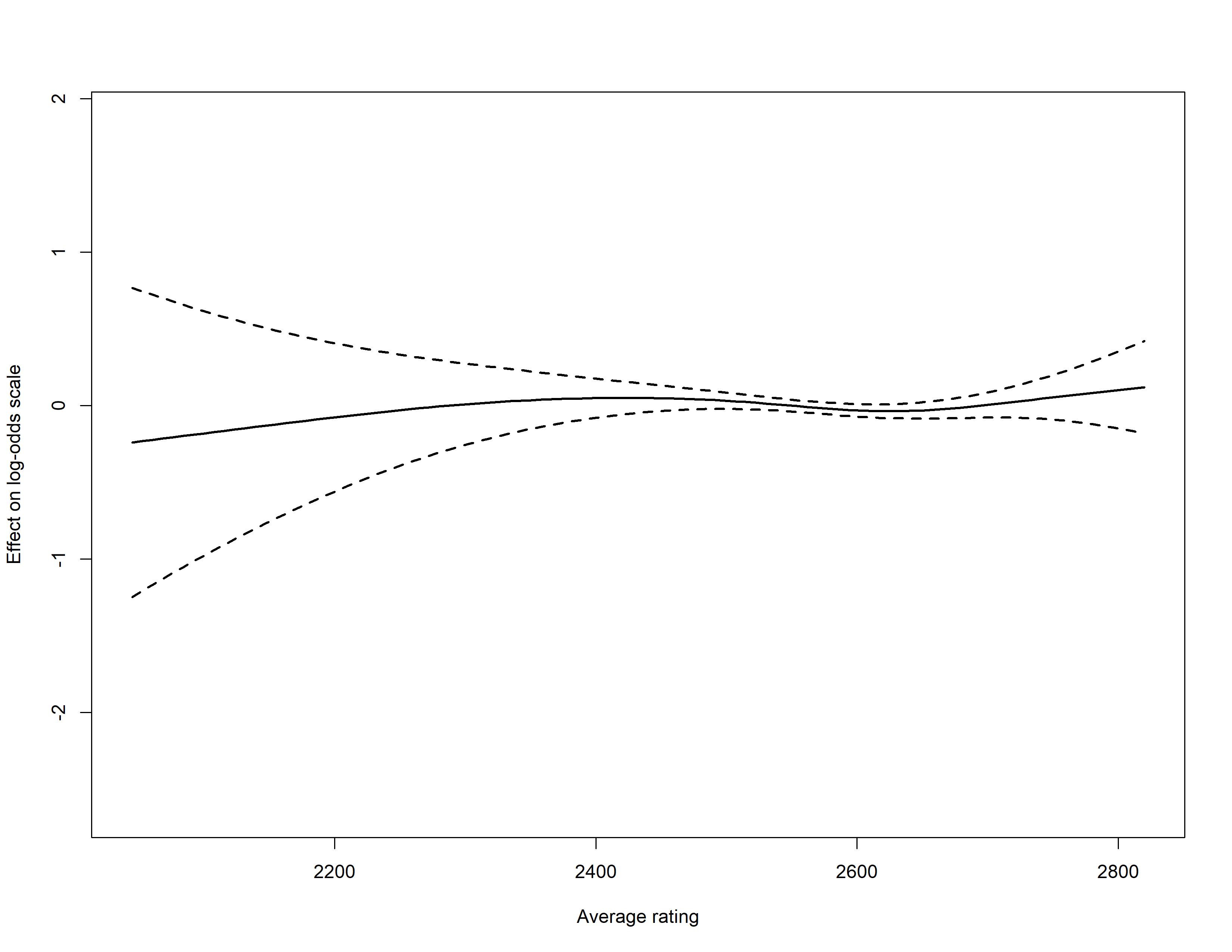}
}
\caption{Log-odds of white winning as a function of the average rating 
between two  players, controlling for rating difference.
The dashed lines are 95\% pointwise confidence limits.
Only decisive games were included in this analysis.}
\label{fig:decis2}
\end{figure}

We also examined whether the probability of a win for white depends on the average
rating for a player pair, controlling for the within-pair rating difference.
To do so, we restricted our attention to the subset of 11,802 games that had a
decisive result (i.e., wins and losses only), and fit a generalized additive model
with a logit link
for the probability that white wins 
as the additive contribution of
a cubic smoothing spline function of the within-pair rating difference and
a cubic smoothing spline function of the within-pair average rating.
Figure~\ref{fig:decis2} displays the fitted smooth of the within-pair average
rating on the log-odds of white winning.
The results of this analysis are less conclusive, but the right side of the 
figure may suggest a slight increase in the log-odds of white winning as
a function of player strength.
Taken together, 
these empirical observations motivate the development of a model in which both the 
tie probability and the order effect vary as a function of the players’ strengths.

\section{A novel paired comparison model}\label{sec:model}

We propose to extend the model by \citet{david_method_1988} to address the described
deficiencies as follows.
For players $i$ and $j$, let
\begin{equation}
\label{eq:novel}
\begin{aligned}
\bdP(Y_{ij}=1) =  &
\dfrac{
\exp\left(\theta_i + x_{ij}\left( \alpha_0 + \alpha_1 \cdot \frac{\theta_i+\theta_j}{2} \right) /4\right)
}{D_{ij}} \\*[12pt]
\bdP(Y_{ij}=0) =  &
\dfrac{
\exp\left(\theta_j - x_{ij}\left( \alpha_0 + \alpha_1 \cdot \frac{\theta_i+\theta_j}{2} \right) /4\right)
}{D_{ij}} \\*[12pt]
\bdP(Y_{ij}=\frac{1}{2}) =  &
\dfrac{
\exp\left( \beta_0 + (1+\beta_1) \cdot \frac{\theta_i+\theta_j}{2} \right)
}{D_{ij}}
\end{aligned}
\end{equation}
% \begin{eqnarray}
% \lefteqn{\bdP(Y_{ij}=1) = }  \nonumber \\
% &&
% \dfrac{
% \exp\left(\theta_i + x_{ij}\left( \alpha_0 + \alpha_1 \cdot \frac{\theta_i+\theta_j}{2} \right) /4\right)
% }{
% \exp\left(\theta_i + x_{ij}\left( \alpha_0 + \alpha_1 \cdot \frac{\theta_i+\theta_j}{2} \right) /4\right) +
% \exp\left(\theta_j - x_{ij}\left( \alpha_0 + \alpha_1 \cdot \frac{\theta_i+\theta_j}{2} \right) /4\right) +
% \exp\left( \beta_0 + (1+\beta_1) \cdot \frac{\theta_i+\theta_j}{2} \right)
% } \nonumber \\*[20pt] 
% \lefteqn{\bdP(Y_{ij}=0) = } \nonumber \\
% &&\dfrac{
% \exp\left(\theta_j - x_{ij}\left( \alpha_0 + \alpha_1 \cdot \frac{\theta_i+\theta_j}{2} \right) /4\right)
% }{
% \exp\left(\theta_i + x_{ij}\left( \alpha_0 + \alpha_1 \cdot \frac{\theta_i+\theta_j}{2} \right) /4\right) +
% \exp\left(\theta_j - x_{ij}\left( \alpha_0 + \alpha_1 \cdot \frac{\theta_i+\theta_j}{2} \right) /4\right) +
% \exp\left( \beta_0 + (1+\beta_1) \cdot \frac{\theta_i+\theta_j}{2} \right)
% } \nonumber \\*[20pt] 
% \lefteqn{\bdP(Y_{ij}=\frac{1}{2}) = } \nonumber \\
% &&\dfrac{
% \exp\left( \beta_0 + (1+\beta_1) \cdot \frac{\theta_i+\theta_j}{2} \right)
% }{
% \exp\left(\theta_i + x_{ij}\left( \alpha_0 + \alpha_1 \cdot \frac{\theta_i+\theta_j}{2} \right) /4\right) +
% \exp\left(\theta_j - x_{ij}\left( \alpha_0 + \alpha_1 \cdot \frac{\theta_i+\theta_j}{2} \right) /4\right) +
% \exp\left( \beta_0 + (1+\beta_1) \cdot \frac{\theta_i+\theta_j}{2} \right)
% } \nonumber \\
% \label{eq:novel}
% \end{eqnarray}
where $\theta_i$ and $\theta_j$ are the strengths of players $i$ and $j$, 
$x_{ij}=1$ if player $i$ plays white and $-1$ if playing black,
$\alpha_0$, $\alpha_1$, $\beta_0$ and $\beta_1$ are parameters of the model
shared across all player pairs, and
\begin{eqnarray}\label{eq:denom}
D_{ij} &=&
   \exp\left(\theta_i + x_{ij}\left( \alpha_0 + \alpha_1 \cdot \frac{\theta_i+\theta_j}{2} \right) /4\right) +
\exp\left(\theta_j - x_{ij}\left( \alpha_0 + \alpha_1 \cdot \frac{\theta_i+\theta_j}{2} \right) /4\right) \nonumber \\
&&+
\exp\left( \beta_0 + (1+\beta_1) \cdot \frac{\theta_i+\theta_j}{2} \right),
\end{eqnarray}
which are the sum of the numerators of~(\ref{eq:novel}).

Compared to David's model in~(\ref{eq:david}), the order effect parameter
$\alpha$ is replaced by $\alpha_0 + \alpha_1(\theta_i+\theta_j)/2$.
Thus, the proposed model assumes an order effect that may increase or decrease as a function
of the average strength of the player pair.
For $\alpha_1=0$, the order effect is treated the same as the 
model by \citet{david_method_1988}.
When $\alpha_1>0$, the probability that white wins increases 
as a function of the average strength of the two competitors,
controlling for the strength difference.

Similarly, 
letting $\beta_0=\log \nu$,
the inclusion of $\nu\cdot\exp((\theta_i+\theta_j)/2)$ in~(\ref{eq:david})
is replaced by 
\[
\nu \cdot \exp((1+\beta_1)(\theta_i+\theta_j)/2) = 
\exp(\beta_0+(1+\beta_1)(\theta_i+\theta_j)/2) .
\]
If $\beta_1=0$, the probability of a draw does not depend on the 
average strength of the player pair, and when both $\alpha_1=0$ and
$\beta_1=0$, the model in~(\ref{eq:novel}) exactly becomes the
\citet{david_method_1988} model in~(\ref{eq:david}).
Our proposed model also preserves IIA 
acknowledging the inclusion of an order effect;
that is, the parameters $\beta_0$ and $\beta_1$ are not relevant
when considering the ratio of the probability of a win to a loss.

\begin{figure}[ht]
\centerline{
\includegraphics[width=1.1\textwidth]{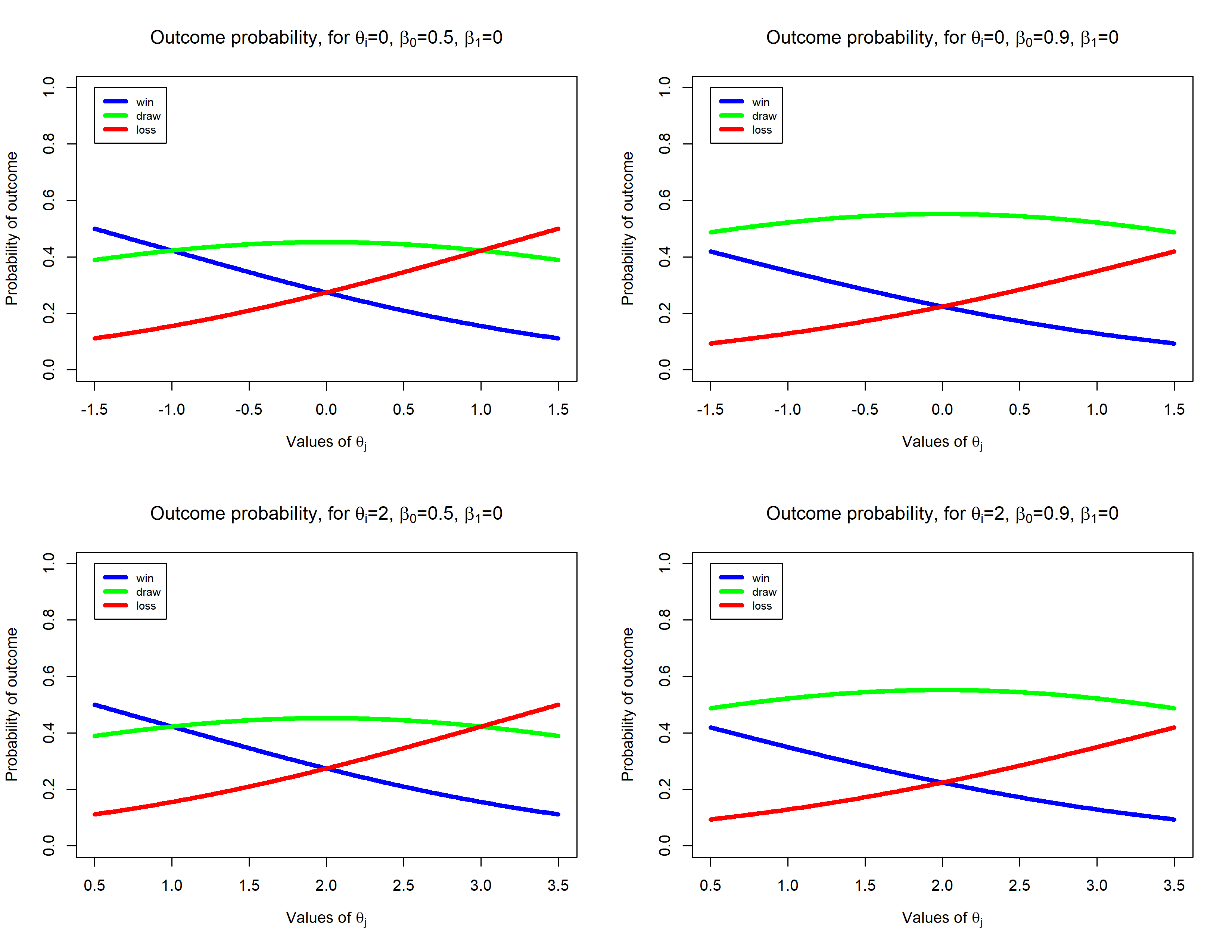}
}
\caption{
Each panel shows outcomes probabilities for a fixed $\theta_i$
(0 on the top row, 2 in the bottom row), across values of $\theta_j$.
Columns correspond to $\beta_0=0.5$ (left) and $\beta_0=0.9$ (right).
Graphs assume $\beta_1=0$, and no order effect, i.e., $\alpha_0=\alpha_1=0$.
}
\label{fig:beta1-0}
\end{figure}

\begin{figure}[ht]
\centerline{
\includegraphics[width=1.1\textwidth]{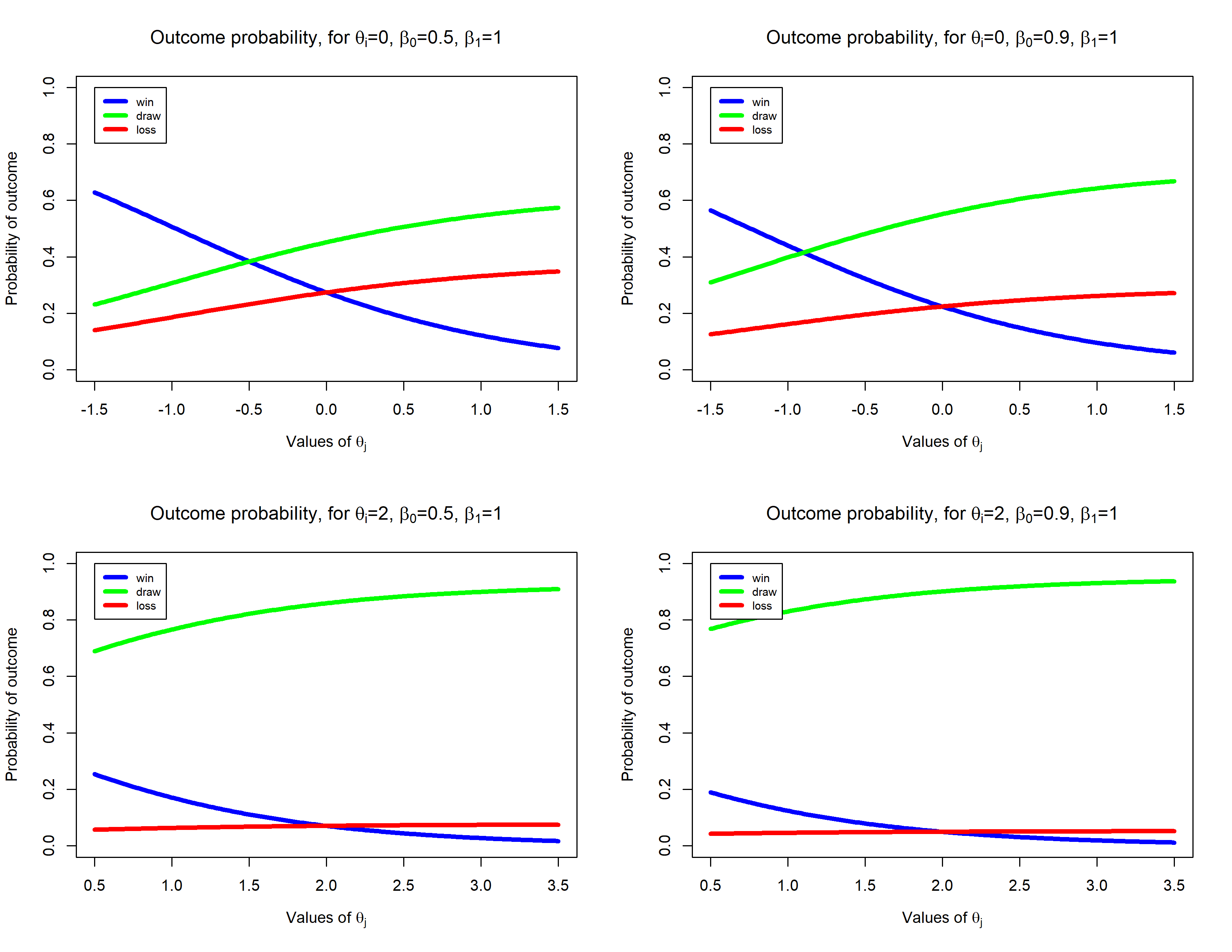}
}
\caption{
Each panel shows outcomes probabilities for a fixed $\theta_i$
(0 on the top row, 2 in the bottom row), across values of $\theta_j$.
Columns correspond to $\beta_0=0.5$ (left) and $\beta_0=0.9$ (right).
Graphs assume $\beta_1=1$, and no order effect, i.e., $\alpha_0=\alpha_1=0$.
}
\label{fig:beta1-1}
\end{figure}

The effect of the choice of $\beta_1>0$ can be understood through 
the probability graphs in Figures~\ref{fig:beta1-0}
and~\ref{fig:beta1-1}.
The four graphs in Figure~\ref{fig:beta1-0} illustrate how outcome probabilities 
vary with the opponent's strength $\theta_j$ when the focal player's 
strength $\theta_i$ is held fixed and the tie parameter $\beta_1$ is set to 0. 
In this case, the probability of a draw 
is constant with respect to the average strength of the players. 
As a result, the draw curves in all four panels are relatively flat and symmetric, 
peaking when $\theta_i=\theta_j$. 
Increasing $\beta_0$ from 0.5 to 0.9 (left vs. right panels) uniformly raises 
the draw probability across all values of $\theta_j$, but does not change 
its shape. 
This behavior contrasts with the graphs in Figure~\ref{fig:beta1-1},
where $\beta_1=1$, allowing the draw probability to increase with the 
average player strength. 
In those graphs, the draw curves rise sharply when both players are strong 
(particularly evident when $\theta_i=2$), and the draw probability dominates 
over probabilities of a win or loss. 
Thus, introducing a positive $\beta_1$ produces outcome dynamics in which 
stronger players are more likely to draw when evenly matched, 
better reflecting empirical patterns 
in games like chess.

Likelihood-based inference for the model in~(\ref{eq:novel}) is straightforward.
The multinomial log-likelihood function is given by
\begin{equation}
\ell(\alpha_0, \alpha_1, \beta_0, \beta_1, \theta_1,\ldots, \theta_n | \bdy)
=
\sum_{k=1}^K
I[y_{k.ij}=1]\log p_{ij1}
+
I[y_{k.ij}=0]\log p_{ij0}
+
I[y_{k.ij}=0.5]\log p_{ijd},
\end{equation}
where $y_{k.ij}$ is the outcome of game $k$, $k=1,\ldots, K$, played
between $i$ and $j$, 
$p_{ij1}$, $p_{ij0}$ and $p_{ijd}$ are the probabilities of a win,
loss and draw, respectively, as defined in Equation~(\ref{eq:novel}),
and $I[y_{k.ij}=z]$ is an indicator of whether game $k$ resulted
in outcome $z$.
Maximum likelihood estimation (MLE) 
for the model may proceed by the method of alternating maximization
\citep{Bezdek2002Some, Ma2015Alternating} in the following manner.
First, the parameters may be divided into two disjoint groups;
the strength parameters $\bdtheta= (\theta_1,\ldots, \theta_K)$,
and the non-strength parameters $\bdgamma = (\alpha_0, \alpha_1, \beta_0, \beta_1)$.
At the first iteration,  set $\bdtheta=0$.
Now with $\bdtheta$ fixed, conditionally maximize log-likelihood
with respect to $\bdgamma$. 
This can be
performed through the Newton-Raphson algorithm 
recognizing that the log-likelihood of $\bdgamma$ is
a standard multinomial logit regression.
Then, conditional on the estimated $\bdgamma$, optimize
the log-likelihood with respect to the $\bdtheta$, which again is
a standard multinomial logit model fitting problem 
with the additional requirement of a linear identifiability constraint
on the $\theta_k$ (e.g., $\sum_{k=1}^K\theta_k=0$)
that can use
standard numerical methods like the Newton-Raphson algorithm.
This conditional alternating optimization proceeds until further changes
in the iterations are below a preset tolerance level.

Likewise, inference for our model
can be embedded in a Bayesian framework, which,
when assuming a proper prior distribution,
can particularly help in situations where players have perfect
scores (all wins or all losses) in which case the strength parameters
have MLEs that are not finite.
Another advantage of the Bayesian framework for chess applications,
in particular, is that players often have pre-tournament ratings
prior to competing, so that a prior distribution centered at a measure
of a player's pre-tournament strength can provide more precise inferences.
We demonstrate such an example in Section~\ref{sec:application}.

For a Bayesian setting, a natural choice of a prior density is
the product of independent normal densities of the $\theta_k$, $k=1,\ldots, K$,
and of $\alpha_0$, $\alpha_1$, $\beta_0$, and $\beta_1$.
In particular, for the strength parameters, we may assume for each $k$
\begin{equation}\label{eq:thetaprior}
    \theta_k \sim \N(\mu_k, \sigma_k^2),
\end{equation}
with prior mean $\mu_k$ and $\sigma_k^2$ pre-specified.
In settings where player $k$ has had past results that have led to
a pre-tournament measure of strength, the normal prior distribution for
$\theta_k$ may be informative with a moderate-sized prior variance.
Without prior information, choosing $\mu_k=0$ may be reasonable, along
with a large prior variance $\sigma_k^2$ that reflects initial uncertainty.

Computation for model fitting in a Bayesian setting can proceed using
Markov chain Monte Carlo (MCMC) simulation from the posterior distribution
via the Gibbs sampler.
Much like the approach of alternating conditional optimization for the MLE,
the MCMC algorithm can alternately sample from the conditional posterior
distribution of $\bdgamma$ given $\bdtheta$, 
and then $\bdtheta$ given  $\bdgamma$.
Each instance is equivalent to sampling from the conditional posterior
distribution of multinomial logit linear coefficients with a normal
prior distribution.
MCMC sampling from such models 
has been developed in \citet{Fruhwirth-Schnatter2007Auxiliary}
and \citet{Fruhwirth-Schnatter2010Data}.
Also, because the model involves a straightforward specification of
multinomial probabilities conditional,
the model can be analyzed in commonly used Bayesian software that
implements MCMC simulation from the posterior distribution.
Model inferences can be summarized based on the simulated samples 
from the posterior distribution upon convergence of the MCMC sampler.

% \section{Inference} \label{sec:inference}

\section{Application to chess game outcomes} \label{sec:application}

% \subsection{US Chess Open: 2006-2019}\label{subsec:usopen}

We demonstrate the application of our modeling framework to 
games results from 
the US Chess Open tournaments from 2006 to 2019. 
The data were obtained in part from the tournament results
posted on the US Chess website (searching for ``US Open'' at\\
https://www.uschess.org/datapage/event-search.php)
as well as directly from the tournament organizers.
The starting year (2006) was the first to our knowledge
in which the US Open collected information on color assignment.
The final year (2019) was the year
before the Covid pandemic started and halted in-person tournament 
competition.
The US Open is an annual tournament taking place during July/August
which attracts participants from all over the United States and internationally.
These tournaments are well-suited to test our modeling framework 
because each participant typically plays 9 games, once per day, 
against distinct opponents.
The format of the tournament is a Swiss System 
\citep{just2019official}
% {\color{red}(citation here)}
in which players are paired against each other with similar
cumulative results from the previous rounds.
This format involves more games per player than usual weekend tournament
formats, which allows for more information about player abilities
than other tournaments.
Also, this format typically pairs players of widely ranging strengths
against each other, usually in the early rounds of the tournament, and
then pairing players of similar skill against each other later in the
tournament.
Thus, in the US Open tournaments, we have more game results per player
to inform whether the probability of a draw and the impact of playing white
depends on player strength.
The information we recorded for each game across the 14 tournaments includes
the two players' US Chess IDs, which player had white, and the chess rating
of each player (if they had one).

Table~\ref{tbl:descrip} summarizes key features of each US Open.
\begin{table}
\centerline{
  \begin{tabular}{r||r||r|r||r||r||}
  US Open & Number  & Number of  & Number of & Total Number & Avg \# of Games \\
  Year    & of Games & Rated Players & Unrated Players & of Players & per Player\\ \hline\hline
2006 & 2182 & 529 & 14 & 543 & 8.037\\ \hline
2007 & 1675 & 407 & 7 & 414 & 8.092\\ \hline
2008 & 1576 & 375 & 4 & 379 & 8.317\\ \hline
2009 & 1888 & 450 & 6 & 456 & 8.281\\ \hline
2010 & 1906 & 469 & 5 & 474 & 8.042\\ \hline
2011 & 1497 & 363 & 4 & 367 & 8.158\\ \hline
2012 & 2191 & 506 & 11 & 517 & 8.476\\ \hline
2013 & 2139 & 516 & 3 & 519 & 8.243\\ \hline
2014 & 1580 & 385 & 1 & 386 & 8.187\\ \hline
2015 & 2013 & 487 & 4 & 491 & 8.200\\ \hline
2016 & 1587 & 389 & 0 & 389 & 8.159\\ \hline
2017 & 1636 & 391 & 5 & 396 & 8.263\\ \hline
2018 & 1568 & 388 & 1 & 389 & 8.062\\ \hline
2019 & 1450 & 350 & 5 & 355 & 8.169\\ \hline \hline
Total & 24888 & 6005 & 70 & 6075 & 8.194\\ \hline\hline
  \end{tabular}
  }
    \caption{Number of games and number of players in each US Chess Open, by year.}
    \label{tbl:descrip}
\end{table}
% year, # total games, avg # games per player, # rated players, # unrated players, total # players
While these tournaments are 9 rounds long, some players do not play all
nine rounds due to byes and forfeits, as well as other idiosyncratic reasons
such as playing abbreviated tournament schedules.
Most players enter the US Open with a US Chess rating
\citep{glickman2024uscf},
which has been computed based on previous tournament results,
but some players compete in the US Open as unrated players.
On average, 429 players have competed in the US Opens from 2006 to 2019,
with a total of 6005 players over the 14 years.
It is worth pointing out that many players have competed in multiple US Opens,
so in fact there are fewer than  6005 players total.
However, from a modeling perspective, we treat players who have competed
in multiple years as distinct players.
This is because players' strengths can and do vary over time, and 
accounting for a player's change in ability over time is beyond the
scope of this work.
By viewing players competing in different years as distinct players,
our inferences will be more conservative than had we explicitly accounted for
the connection over time.

We considered six different models, all of which are special cases of 
our proposed model in Equation~(\ref{eq:novel}).
They are summarized in Table~\ref{tbl:models}.
\begin{table} 
    \begin{tabular}{r|l}
    Model & Description \\ \hline\hline
    Model 1 & Full model \\ \hline
    Model 2 & Model with no color effect ($\alpha_0=\alpha_1=0$)\\ \hline
    Model 3 & Model with constant color effect ($\alpha_1=0$)\\ \hline
    Model 4 & Full model, non-varying draw probability ($\beta_1=0$)\\ \hline
    Model 5 & Model with no color effect, non-varying draw probability 
($\alpha_0=\alpha_1=\beta_1=0$)\\ \hline
    Model 6 & Model with constant color effect and non-varying draw probability
($\alpha_1=\beta_1=0$) \\ \hline \hline
    \end{tabular}
    \caption{Model variants explored.}
\label{tbl:models}
\end{table}
Model~1 is the most general model, and is the one specified in~(\ref{eq:novel}).
Models~2 through~5 are various special cases of the general model,
with certain parameters set to 0,
and Model~6 is the model by \citet{david_method_1988}, which itself is
a special case of Model~1.

In addition to the different variants of the probability model for game
outcomes, we also considered two different prior distributions for the
player strength parameters.
The first approach was to assume an informative prior distribution for
player strengths depending on whether the player had a pre-tournament
rating.
For all $i$, we assumed
\begin{equation}
\label{eq:prior-inf}
  \theta_i \sim  \left\{
  \begin{array}{rl}
    \N(\mu_i, \sigma^2) & 
      \mbox{if player $i$ had a pre-tournament rating $\mu_i$}\\*[8pt]
    \N(\mu_{\text{miss}}, \sigma_{\text{miss}}^2) &
      \mbox{if player $i$ was unrated,}
  \end{array}
  \right.
\end{equation}
where $\mu_i = (R_i - 1500)/(400/\log 10)$, with $R_i$ the pre-tournament
US Chess rating measured on the Elo rating scale.
This assumption reflects that player strengths with pre-tournament ratings
have {\em a priori} similar precision about their true strengths.
For players who were unrated, we assumed an exchangeable normal prior distribution
with a variance different from the rated players.
The second approach to a prior distribution was to assume an exchangeable
normal prior distribution for all players, and ignore their pre-tournament
ratings if they had one.
Thus, we assumed for all $i$,
\begin{equation}
\label{eq:prior-exch}
 \theta_i \sim \N(0, \sigma^2),
\end{equation}
that is, a normal prior for all players' strengths with a common mean of zero
and a common variance.
Because our model is invariant to additive shifts to the strength parameters,
assuming a zero mean is no less meaningful than other choices.

In total, we therefore fit a total of 12 models; the six listed in 
Table~\ref{tbl:models} with an informative prior distribution on 
player strengths, and
the six models with an exchangeable noninformative prior on the player strengths.
The specification of the models were completed by assuming vague prior
distribution components on the non-strength parameters and the variance
parameters.
For $\alpha_0$, $\alpha_1$, $\beta_0$ and $\beta_1$, we assumed 
normal indepedent prior distributions with mean 0 and variance 100.
The variance parameters $\sigma^2$ and $\sigma_{\text{miss}}^2$ were
assumed to follow inverse-Gamma distributions with parameters
0.01, and 0.1, corresponding to a mean and variance that are not finite,
and thus noninformative.
With a large number of observations informing these variance parameters,
concerns with the impact of assuming a diffuse prior distribution were minimal.

We implemented our model-fitting using
JAGS \citep{plummer_jags_2003} within R via the R2Jags library
% \citep{plummer2022rjags}.  
\citep{su2024package}.
Each model was fit using 3 parallel chains with dispersed starting values.
The chains were run for 20,000 iterations, treating the first 10,000
iterations as burn-in and keeping the final 10,000 iterations for
model summaries.
To reduce auto-correlation in the simulated draws, we kept only every
fifth value for each chain.
This resulted in a total of 6,000 simulated values for each
parameter set.
Convergence of the chains were assessed via the $\hat{R}$ statistic
\citep{gelman1992inference} for the non-strength parameters
as implemented in JAGS.
After 20,000 iterations, all $\hat{R}$ values were less than 1.01,
indicating that the chains had mixed well after the burn-in period.

Table~\ref{tbl:dic} summarizes the fits of the model using the 
deviance information criterion (DIC)
statistic \citep{spiegelhalter_bayesian_2002}.
\begin{table}
    \centerline{
    \begin{tabular}{r|r|r|}
          & Informative & Noninformative\\
    Model & Prior & Prior \\ \hline
    Model 1 & 43821.32 & 53686.01 \\
    Model 2 & 43956.22 & 54253.84 \\
    Model 3 & 44086.10 & 54771.51 \\
    Model 4 & 44552.77 & 54593.37 \\
    Model 5 & 44713.30 & 54120.95 \\
    Model 6 & 44658.40 & 54926.64 \\ \hline
    \end{tabular}
    }
    \caption{DIC values for each of the 12 models.}
    \label{tbl:dic}
\end{table}
Lower values of the DIC indicate a better fit,
and it is common practice when comparing models that
differences in DIC 
greater than 2-3 suggest support for the model with the lower DIC
\citep{wheeler2010assessing}.
Using this guideline, the full model (Model 1) with an informative
prior distribution based on pre-tournament ratings,
fits the US Open tournament data substantially better 
than any other variant of our model, including the \citet{david_method_1988}
model (Model 6).
Among the models with a noninformative prior distribution,
the full model also substantially outperforms the other models,
but underperforms relative to all of the models with an informative prior
distribution on the strength parameters.

We summarize inferences for the non-strength model parameters
for our best-fitting model in Table~\ref{tbl:inferences}.
All model parameters are estimated precisely, with 95\% central posterior intervals 
closely concentrated around their respective posterior means.
Since most of the posterior mass of $\alpha_0$ lies well above zero, 
the advantage of playing white for players of average ability 
(i.e., when $(\theta_i+\theta_j)/2 \approx 0$) appears substantial.
For two average players, an estimate of $\alpha_0 = 0.363$ implies 
win-to-loss odds of $\exp(0.363/2) = 1.2$ in favor of white. 
This corresponds to white and black winning probabilities of 0.545 and 0.455, 
respectively, among decisive games between evenly matched players.
\begin{table}[t]
    \centerline{
    \begin{tabular}{r|r|r|}
          & Posterior & 95\% Central \\
    Parameter & Mean & Posterior Interval\\ \hline
    $\alpha_0$ & 0.363 & (0.289, 0.434) \\
    $\alpha_1$ & 0.037 & (0.000, 0.074) \\
    $\beta_0$ & --0.471 & (--0.505, --0.437) \\
    $\beta_1$ & 0.120 & (0.103, 0.138) \\
    $\sigma$ & 0.645 & (0.604, 0.687) \\
    $\mu_{\text{miss}}$ & --3.399 & (--4.110, --2.703) \\
    $\sigma_{\text{miss}}$ & 2.636 & (2.085, 3.310) \\ \hline
    \end{tabular}
    }
    \caption{Summaries of parameters from best-fitting model.
    Posterior means and 95\% central posterior intervals are computed based on 
    the 6,000 simulated values from posterior simulations.}
    \label{tbl:inferences}
\end{table}
The value of $\alpha_1$ being inferred to be positive indicates that
the advantage to playing white increases slightly as a function of the players' strengths.
While two evenly matched players of average strength have white-to-black win odds of 
approximately 1.2, two evenly matched strong players 
(with $\theta_i = \theta_j = 2$) have estimated odds of 
$\exp((0.363 + 2 \times 0.037)/2) = 1.244$ according to the model. 
This corresponds to a white win probability of 0.554 among decisive games, 
slightly higher than the 0.545 probability for average-strength players.
The estimate of $\beta_0$ of $-0.471$ indicates that for evenly matched players of
average strength, assuming no advantage to white (setting $\alpha_0=\alpha_1=0$), 
the probability of a draw is
$\exp(-0.471)/(1 + 1 + \exp(-0.471)) = 0.238$.
Given that $\beta_1$ is inferred to be larger than 0, the analogous probability
of a draw between two strong players (with $\theta_i=\theta_j=2$, 
and $\alpha_0=\alpha_1=0$ corresponding to no advantage to white) is 
$\exp(-0.471 + (1+0.120))/(1 + 1 + \exp(-0.471 + (1+0.120))) = 0.489$, 
a much higher probability.
The model fit also results in a small estimated value of $\sigma$, the standard
deviation of the prior distribution of a player's strength centered at their
pre-tournament estimate.
This is notably lower than the estimate of $\sigma_{\text{miss}}$, the prior 
standard deviation of a player's strength when no pre-tournament rating was
provided.
Similarly, the estimate of $\mu_{\text{miss}} = -3.399$ suggests that unrated players were, 
on average, substantially weaker than their rated counterparts.

\begin{figure}[ht]
\centerline{
\includegraphics[width=1.1\textwidth]{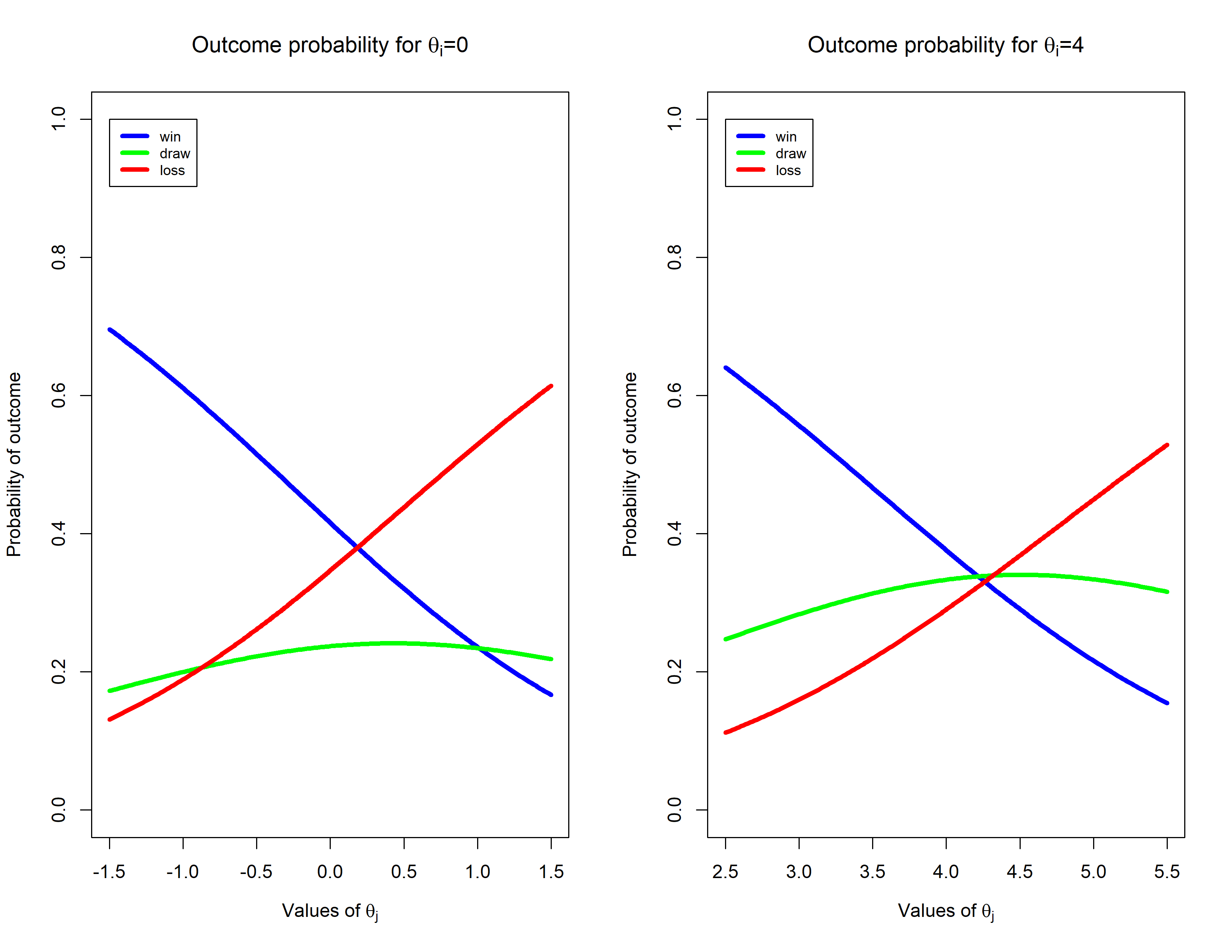}
}
\caption{
Each panel shows outcomes probabilities for a fixed $\theta_i$
(0 on the left, 4 on the right), across values of $\theta_j$.
The focal player is assumed to be playing as white.
The values of $\alpha_0$, $\alpha_1$, $\beta_0$ and $\beta_1$
are the posterior means given in Table~\ref{tbl:inferences}.
}
\label{fig:modelfit}
\end{figure}

The two panels in Figure~\ref{fig:modelfit} display the predicted probabilities of 
win (blue), draw (green), and loss (red) outcomes for a player with strength 
$\theta_i = 0$ (left panel) and $\theta_i = 4$ (right panel), as a function of the 
opponent’s strength parameter $\theta_j$, using the posterior mean estimates of the 
model parameters. 
In both panels, as $\theta_j$ increases, the win probability decreases and the 
loss probability increases, reflecting the effect of facing a stronger opponent. 
% However, there are notable contrasts between the two panels. 
In the left panel ($\theta_i = 0$), the win probability begins high (around 0.7) 
and declines to approximately 0.2, while the loss probability starts low and rises 
past 0.6. 
The higher win probability at $\theta_j=-1.5$ compared to the lower loss probability at
$\theta_j=1.5$ recognizes that the advantage of playing white.
The draw probability remains relatively low throughout and exhibits only a mild peak 
just to the right of $\theta_j = \theta_i$, which again acknowledges the advantage
to playing white when $\theta_j=\theta_i$ exactly.

In contrast, in the right panel ($\theta_i = 4$), 
the focal player is much stronger. 
Importantly, the draw probability in the right panel is uniformly higher than in the 
left panel and exhibits a more pronounced peak to the right of $\theta_i = \theta_j$, 
illustrating the model’s feature that draws become more likely when stronger players 
face each other. 
Thus, the comparison reveals that the likelihood of a draw increases not only when 
the players are closely matched in strength, but also as their absolute level of 
strength increases.

% \subsection{Vienna 1898 Chess Tournament}  \label{subsec:vienna}

\section{Discussion} \label{sec:discuss}

This paper developed a novel extension of the Bradley–Terry paired comparison framework 
that jointly accounts for tie probabilities and order effects, allowing both to vary 
with the average strength of the competing players. 
This modeling approach was motivated by empirical evidence, particularly from chess, 
indicating that stronger players tend to draw more frequently and may be more adept at 
leveraging structural advantages such as playing white. 
The proposed model replaces traditional fixed order-effect and tie parameters 
with strength-dependent versions, yielding a richer and more realistic representation 
of competitive outcomes. 
Application of the model to US Chess Open tournaments from 2006 to 2019 demonstrated 
its empirical relevance, with clear improvements in predictive performance compared 
to traditional formulations.

While the model represents a substantial generalization of prior approaches, it is not 
without limitations. 
The linear dependence of tie and order effects on average player strength, 
while interpretable and parsimonious, may not fully capture more complex or 
nonlinear relationships that can arise in practice. 
A natural extension of this work would involve replacing the linear effect of
the within-pair average strength with 
smooth functions, such as monotone splines or kernel-based transformations, to better model 
the complex behavior in the tie frequency or order effect as a function of strength.
Additionally, the current framework treats each player-year as independent, 
ignoring potential continuity in player ability over time. 
A dynamic formulation, such as a state-space model where player strengths evolve 
longitudinally, 
would allow for tracking changes in player strength across tournaments, 
enabling coherent modeling of repeated participation.

The modeling innovations presented here are particularly well-suited to sports settings 
where ties are common and home-field advantage plays a substantial role. 
Sports such as soccer, hockey, and boxing often feature both tied outcomes and 
varying contextual advantages, and the ability to model these features as functions 
of player or team strength enhances both predictive power and interpretability. 
By capturing how tie probabilities and order effects interact with ability, 
the framework presented here offers a more nuanced tool for understanding 
head-to-head competition and can help inform strategic decisions in 
tournament design, ranking systems, and player evaluation.

\section*{Acknowledgments}
Thanks to Carlos Pena-Lobel for developing software to convert 
US Chess tournament crosstables into an analyzable format used in 
this work.

\addtolength{\baselineskip}{-12pt}
\bibliographystyle{agsm}
\bibliography{refs}

% \begin{appendices}
% \section{Proof of Theorem~\ref{theorem:convergence}}
% \end{appendices}
\end{document}